\documentclass[twocolumn,prd,showpacs,amsmath,amssymb]{revtex4}

\begin{document}
\title{ Operational definition of (brane induced) space-time \\and constraints on the fundamental parameters
\footnote{Extended version of the talk given at ICTP.}}

\author{ Michael~Maziashvili}
\email{maziashvili@hepi.edu.ge} \affiliation{ Andronikashvili
Institute of Physics, 6 Tamarashvili St., Tbilisi 0177, Georgia }

\begin{abstract}
First we contemplate the operational definition of space-time in
four dimensions in light of basic principles of quantum mechanics
and general relativity and consider some of its phenomenological
consequences. The quantum gravitational fluctuations of the
background metric that comes through the operational definition of
space-time are controlled by the Planck scale and are therefore
strongly suppressed. Then we extend our analysis to the braneworld
setup with low fundamental scale of gravity. It is observed that
in this case the quantum gravitational fluctuations on the brane
may become unacceptably large. The magnification of fluctuations
is not linked directly to the low quantum gravity scale but rather
to the higher-dimensional modification of Newton's inverse square
law at relatively large distances. For models with compact extra
dimensions the shape modulus of extra space can be used as a most
natural and safe stabilization mechanism against these
fluctuations.

\end{abstract}



\maketitle

\subsection*{Introduction}

From the inception of quantum mechanics the physical quantities
are usually understood to be observable, that is, they should be
specified in terms of real or {\tt Gedanken} measurements
performed by well-prescribed measuring procedures. The concept of
measurement has proved to be a fundamental notion for revealing
the genuine nature of physical reality \cite{Heisenberg}.
Space-time representing a frame in which everything takes place is
one of the most fundamental concepts in physics. The importance of
operational definition of physical quantities gives a strong
motivation for a critical view how one actually measures the
space-time geometry \cite{SW, Mead}. The first natural question in
this way is to understand to what maximal precision can we mark a
point in space by placing there a test particle. Throughout this
paper we will use system of units $\hbar = c = 1$. In the
framework of quantum field theory a quantum takes up at least a
volume, $\delta x^3$, defined by its Compton wavelength $\delta x
\gtrsim 1/m$. Not to collapse into a black hole, general
relativity insists the quantum on taking up a finite amount of
room defined by its gravitational radius $\delta x \gtrsim
l_P^2m$. Combining together both quantum mechanical and general
relativistic requirements one finds
\begin{equation}\label{abslimit} \delta x \gtrsim
\mbox{max}(m^{-1},~l_P^2m)~.\end{equation} From this equation one
sees that a quantum occupies at least the volume $ \sim l_P^3 $.
Therefore in the operational sense the point can not be marked to
a better accuracy than $ \sim l_P^3 $. As any measurement we can
perform (real or {\tt Gedanken}) is based on the using of quanta,
from Eq.(\ref{abslimit}) one infers that we can never probe a
length to a better accuracy than $\sim l_P$. Since our
understanding of time is tightly related to the periodic motion
along some length scale, this result implies in general an
impossibility of space-time distance measurement to a better
accuracy than $\sim l_P$. This point of view was carefully
elaborated in \cite{Mead}. This apparently trivial conclusion
encountered serious bias when it was originally suggested by Mead
\cite{MW}. Starting from the 1980s the operational definition of
space-time attracted considerable continuing interest \cite{Pady,
Maggiore, Ahluwalia, AC, NvD, Sasakura}.

Our fundamental theories of physics involve huge hierarchies
between the energy scales characteristic of gravitation $E_P =
1/\sqrt{G_N} \sim 10^{28}$eV and particle physics $E_{EW} \sim
1$TeV. In the atomic and subatomic world therefore, gravity is so
weak as to be negligible. This is one reason gravity is not
included as part of the Standard Model of particle physics. But
when energy scale approaches the Planck one gravity enters the
game. The question of operational definition of space-time becomes
particulary interesting and important in regard with the
higher-dimensional theories with low quantum scale of gravity
(close to the electroweak scale). First we summarize different
approaches for operational definition of Minkowskian space-time
that enables one to estimate the rate of quantum-gravitational
fluctuations of the background metric. Then we address some of the
implications of these fluctuations. Having discussed the case of
4D space-time, we generalize the operational definition to the
brane induced space-time and consider its phenomenological
consequences.

\subsection*{K\'arolyh\'azy uncertainty relation}

{\bf \emph{Approach} 1. -} For space-time measurement an
unanimously accepted method one can find in almost every textbook
of general relativity consists in using clocks and light signals
\cite{FockLL}. Let us consider a light-clock consisting of a
spherical mirror inside which light is bouncing. That is, a
light-clock counts the number of reflections of a pulse of light
propagating inside a spherical mirror. Therefore the precision of
such a clock is set by the size of the clock. The points between
which distance is measured are marked by the clocks, therefore the
size of the clock $2r_c$ from the very outset manifests itself as
an error in distance measurement. Another source of error is due
to quantum fluctuations of the clocks. Namely denoting the mass of
the clock by $m$ one finds that the clock is characterized with
spread in velocity

\[\delta v = {\delta p \over m} \sim {1 \over m\, r_c}~,\] and
correspondingly during the time $t$ taken by the light signal to
reach the second clock the clock may move the distance $t\delta
v$. The total uncertainty in measuring the lengths scale $l$ takes
the form \[\delta l \gtrsim r_c + {l\over m\,r_c}~. \] Minimizing
this expression with respect to the size of clock one finds
\begin{equation}\label{minwitsize}r_c \simeq \sqrt{l\over m}~~\Rightarrow~~~~\delta l \gtrsim
\sqrt{l\over m}~.\end{equation} By taking the mass of the clock to
be large enough the uncertainty in length measurement can be
reduced but one should pay attention that simultaneously the size
of the clock diminishes and its gravitational radius increases.
The measurement procedure to be possible we should care the size
of the clock not to become smaller than its gravitational radius
to avoid the gravitational collapse of the clock into a black
hole. So that there is an upper bound on the clock mass
\[r_c^{min} \simeq \sqrt{{l\over m_{max}}} \simeq l_P^2
m_{max}~,~~\Rightarrow ~~ m_{max} \simeq {l^{1/3}\over
l_p^{2/3}}~,\] which through the equation (\ref{minwitsize})
determines the minimal unavoidable error in length measurement as
\begin{equation}\label{Karol}\delta l_{min} \simeq l_P^{2/3}l^{1/3}
~.\end{equation} This way of reasoning follows to the papers
\cite{SW, NvD}.

{\bf \emph{Approach} 2. -} One can discuss in a bit different way
as well \cite{Sasakura}. Let us consider the construction of a
coordinate system for a time interval $t$ and with a spatial
fineness $\delta x$ in a Minkowski space-time. Since a clock must
be localized in a region with the size $\delta x$, the clock
inevitably has a momentum of the order $\delta p\sim 1/\delta x$,
obtained from the uncertainty relation of quantum mechanics. Thus
the clock moves with a finite velocity of order $\delta v \sim
1/m\delta x$, where $m$ denotes the mass of the clock. This
implies that the coordinate system will be destroyed by the
quantum effect in a finite period $\delta x/\delta v\sim m(\delta
x)^2 $. This period must be larger than the time interval $t$ of
the coordinate. Hence we obtain
\begin{equation} t\lesssim m(\delta x)^2~. \label{eq:ineqtx}
\end{equation} This gives a lower bound for the clock mass $m$
for given $t$ and $\delta x$. From Eq.(\ref{eq:ineqtx}), we need
clock with a larger mass to construct a finer coordinate system.
However we also have a maximum value of a clock mass, because no
clock should become a black hole. Thus the clock's Schwarzschild
radius should not exceed the localization region of the clock:
\begin{equation} l_P^2m \lesssim \delta x. \label{eq:gmx}
\end{equation} The clock mass can be chosen arbitrary if it
satisfies Eq.(\ref{eq:ineqtx}) and Eq.(\ref{eq:gmx}). Combining
Eqs.(\ref{eq:ineqtx},\,\ref{eq:gmx}) one gets
\begin{equation} l_P^2 t \lesssim (\delta x)^3. \label{eq:txxx}
\end{equation} Taking note that our light-clock having the size $\delta
x$ can not measure the time to a better accuracy than $\delta t =
\delta x$ one arrives at the Eq.(\ref{Karol}).

{\bf \emph{Approach} 3. -} It is instructive to take into account
gravitational time delay of the clock \cite{mazia}. After
introducing the clock the metric takes the form

\[ds^2=\left(1-{2l_P^2m\over r}\right)dt^2- \left(1-{2l_P^2m\over r}\right)^{-1}dr^2-r^2d\Omega^2~.\]
The time measured by this clock is related to the Minkowskian time
as \cite{FockLL} \[t'=\left(1-{2l_P^2m\over r_c}\right)^{1/2}t~.\]
From this expression one sees that the disturbance of the
background metric to be small, the size of the clock should be
much greater than its gravitational radius $r_c \gg 2l_p^2m$.
Under this assumption for gravitational disturbance in time
measurement one finds \[t'=\left(1-{l_P^2m\over r_c}\right)t~.\]
Since we are using light-clock its mass can not be less than
$\pi/r_c$, which by taking into account that the size of the clock
determining its resolution time represents in itself an error
during the time measurement gives \[\delta t =
2r_c+\pi{tt_P^2\over r_c^2}~,\] which after minimization with
respect to $r_c$ leads to the Eq.(\ref{Karol}).

What is common in all of the above approaches is the final result
Eq.(\ref{Karol}). Nevertheless the third approach strongly
discourages to take the optimal size of the clock to be close to
its gravitational radius. The first and second approaches do not
take into account the gravitational time delay of the clock. For
the optimal parameters of the clock in measuring the space-time
distance $l$ one finds
\[r_c \simeq l_P^{2/3}\,l^{1/3}~,~~~~~~~~m \simeq {1\over r_c}~.\]

Eq.(\ref{Karol}) was first obtained by K\'arolyh\'azy in 1966 and
was subsequently analyzed by him and his collaborators in much
details \cite{Karol}.

\subsection*{Field theory view}

Effective quantum field theory with built in IR and UV cutoffs
satisfying the black-hole entropy bound leads to the
Eq.(\ref{Karol}), where $l$ and $\delta l$ play the roles of IR
and UV scales respectively \cite{CKN}. For an
 effective quantum field theory in a box of size
$l$ with UV cutoff $\Lambda$ the entropy $S$ scales as, \[S \sim
 l^3\Lambda^3~.\] That is, the effective quantum field theory
 counts the degrees of freedom simply as the numbers of cells
 $\Lambda^{-3}$ in the box $l^3$. Nevertheless, considerations involving black holes demonstrate
that the maximum entropy in a box of volume $l^3$ grows only as
the area of the box \cite{Bekenstein} \[S_{BH} \simeq
\left({l\over l_P}\right)^2~.\] So that, with respect to the
Bekenstein bound \cite{Bekenstein} the degrees of freedom in the
volume should be counted by the number of surface cells $l_P^2$. A
consistent physical picture can be constructed by imposing a
relationship between
 UV and IR cutoffs \cite{CKN}
\begin{equation}\label{BHbound}l^3 \Lambda^3 \lesssim S_{BH} \simeq
 \left({l\over l_P}\right)^2~.
\end{equation}
Consequently one arrives at the conclusion that the length $l$,
 which serves as an IR cutoff, cannot be chosen
independently of the UV cutoff, and scales as $\Lambda^{-3}$.
Rewriting this relation wholly in length terms, $\delta l \equiv
\Lambda^{-1}$, one arrives at the Eq.(\ref{Karol}). Is it an
accidental coincidence? Indeed not. The relation (\ref{BHbound})
can be simply understood from the Eq.(\ref{Karol}). The IR scale
$l$ can not be given to a better accuracy than $\delta l \simeq
l_P^{2/3}\,l^{1/3}$. Therefore, one can not measure the volume
$l^3$ to a better precision than $\delta l^3 \simeq l_P^2\,l$ and
correspondingly maximal number of cells inside the volume $l^3$
that may make an operational sense is given by $(l/l_P)^2$. Thus
the K\'arolyh\'azy relation implies the black-hole entropy bound
given by Eq.(\ref{BHbound}). These ideas lead to the far reaching
holographic principle for an ultimate unification that may perhaps
be achieved when the basic aspects of quantum theory, particle
theory and general relativity are combined \cite{tHooft}.

\subsection*{Energy density of the fluctuations}

K\'arolyh\'azy uncertainty relation naturally translates into the
metric fluctuations, as if it was possible to measure the metric
precisely one could estimate the length between two points
exactly. As we are dealing with the Minkowskian space-time the
rate of metric fluctuations over a length scale $l$ can be simply
estimated through the Eq.(\ref{Karol}) as \[\delta g_{\mu\nu} \sim
{\delta l \over l} \sim \left({l_P\over l} \right)^{2/3}~.\] We
naturally expect there to be some energy density associated with
the fluctuations. One can use the following simple reasoning for
estimating the energy budget of Minkowski space \cite{Sasakura,
mazia}. With respect to the Eq.(\ref{Karol}) a length scale $t$
can be known with a maximum precision $\delta t$ determining
thereby a minimal detectable cell $\delta t^3 \simeq t_P^2t$ over
a spatial region $t^3$. Such a cell represents a minimal
detectable unit of space-time over a given length scale and if it
has a finite age $t$, its existence due to time energy uncertainty
relation can not be justified with energy smaller then $\sim
t^{-1}$. Hence, having the above relation, Eq.(\ref{Karol}), one
concludes that if the age of the Minkowski space-time is $t$ then
over a spatial region with linear size $t$ (determining the
maximal observable patch) there exists a minimal cell $\delta t^3$
the energy of which due to time-energy uncertainty relation can
not be smaller than
\begin{equation} \label{cellenergy} E_{\delta t^3} \gtrsim t^{-1}~.
\end{equation} Hence, for energy density of metric
fluctuations of Minkowski space one finds
\begin{equation}\label{qenerdensi} \rho \sim {E_{\delta t^3}\over \delta t^3} \sim {
1 \over t_P^2 t^2}~, \end{equation} which for $t\sim H_0^{-1}$
gives the observed value \cite{PerlRiess}\[\rho_0 \sim {H_0^2
\over l_p^2}~.\] The time will lose its physical meaning when
$\delta t \gtrsim t$ which is tantamount to the decreasing of
background energy density, Eq.(\ref{qenerdensi}), below the
$\lesssim t^{-4}$. One can say the existence of this background
energy density assures maximal stability of Minkowski space-time
against the fluctuations as the Eq.(\ref{Karol}) determines
maximal accuracy allowed by the nature.

On the basis of the above arguments one can go further and see
that due to K\'arolyh\'azy relation, the energy $E$ coming from
the time energy uncertainty relation $E\,t \sim 1$ is determined
with the accuracy $\delta E \sim  E \delta t/ t$. Respectively,
one finds that the energy density $\rho = E / \delta t^3$ is
characterized by the fluctuations $\delta \rho = \delta E/ \delta
t^3$ giving
\begin{equation}\label{reldensfluct}{\delta \rho \over \rho}\sim
{\delta t \over t} \sim \left({t_P \over
t}\right)^{2/3}~.\end{equation} The attempts to estimate the
dynamics of dark energy predicted by the K\'arolyh\'azy relation
during the cosmological evolution of the universe and other
cosmological implications can be found in \cite{Agegraphic}.

\subsection*{Experimental signatures}

A question of paramount importance is to estimate the observable
effects induced by the quantum gravitational fluctuations of the
background metric. Metric fluctuations naturally produce the
uncertainties in
 energy-momentum measurements, for the particle with momentum $p$ has the wavelength $\lambda = 2\pi
p^{-1}$ and due to length uncertainty one finds $\delta p =
2\pi\lambda^{-2}\delta\lambda,~ \delta E = pE^{-1}\delta p$. An
interesting idea for detecting the space-time fluctuations was
proposed in \cite{LH}. The theoretical framework put forward in
\cite{LH} to describe the incoherence of light from distant
astronomical sources due Planck scale quantum gravitational
fluctuations of the background metric is as follows. It is assumed
that the light coming from the distant extragalactic sources, the
diffraction/interference images of which are seen through the two
slit telescopes is coherent from the beginning but can accumulate
 appreciable phase incoherence $t\delta\omega$ even for small $\delta\omega$ caused
by the quantum gravitational fluctuations of the background metric
if the length of propagation, $t$, is large enough. So it is
simply understood that the time Dependance of the wave, $t\omega$,
varies due to quantum gravitational fluctuations as $\delta
(t\omega) = \omega\delta t + t\delta\omega$ and because the second
term is dominating it is taken as a main source of phase
incoherence. The condition $t\delta\omega \geq 2\pi$ is understood
as a criterion for incoherence that should lead to the destroy of
the diffraction/interference patterns when the source is viewed
through a telescope. In \cite{RTG} the distance through which the
wave-front recedes when the phase increases by $t\delta\omega$ is
taken as an error in measurement of a length, $t$, by the light
with wavelength $2\pi/\omega$, and due to this length variation an
apparent blurring of distant point sources was estimated. In
\cite{NgvDC} to mitigate the situation the cumulative factor
$t/\lambda$ in phase incoherence
\begin{equation}\label{cumfac}t\delta\omega = \omega\,{t\over
\lambda}\,\delta\lambda~,\end{equation} was replaced (actually in
an ad hoc manner) by $(t/\lambda)^{1/3}$. This reduced
 expression
for the phase incoherence is used in \cite{Stein} as well. Soon
after the
 appearance of the paper \cite{LH} it was noticed in \cite{Coule} that such a naive approach
 overestimates the effect as the
authors of \cite{LH} do not take into account the van Cittert -
Zernike formalism  representing basics of stellar interferometry
\cite{BornWolf}. Actually the rate of this effect is
discouragingly small to be detectable by the stellar
interferometry observations \cite{MazTin}.

Let us emphasize the main points ignored in \cite{LH}, which prove
to be important in estimating the correct rate of the effect.
Light from a real physical source is never strictly monochromatic
but rather quasi-monochromatic, even the sharpest spectral line
has a finite width. In a wave produced by a real source: the
amplitude and phase undergo irregular fluctuations,
 the
rapidity of which depends on the width of spectrum $\delta\omega$.
Such a quasi-monochromatic wave which is usually referred to as a
wave packet is characterized with a mean wave frequency
$\bar{\omega}$, where \begin{equation}\label{wavepacket}{\delta
\omega \over \bar{\omega}} \ll 1 ~.\end{equation} The width
$\delta\omega$ determines duration of the wave packet $\delta t
\simeq \delta\omega^{-1}$, which is an important characteristic
for the interference effect during a superposition of the
quasi-monochromatic beams. Namely, the interference effect to take
place the path difference between quasi-monochromatic beams must
be small than the coherence length $\delta t$. There is an
increment of the wave packet width due to background metric
fluctuations which can be simply estimated as
\[\delta \omega = \bar{\omega}\, {\delta\lambda \over
\bar{\lambda}} \simeq
\bar{\omega}\,\left({l_P\over\bar{\lambda}}\right)^{2/3}~.
\]  A
 wavelength of the light from
stellar objects considered in \cite{LH, RTG, Stein} is in the
region $\bar{\lambda} \simeq \mu$m and correspondingly for the
width increment of a wave packet one finds \[{\delta \omega \over
 \bar{\omega}} \simeq \,
10^{-19}~.\] Such a small increment does not affect neither the
Eq.(\ref{wavepacket}) nor the requirement the path difference
between quasi-monochromatic beams coming from distant stellar
objects to be small than the coherence length $\delta \omega
^{-1}$ \cite{MazTin}. The expression that comes from the van
Cittert - Zernike approach has the form \cite{BornWolf}
\begin{equation}\label{explim} D =
 {0.16\, \bar{\lambda}\, r \over\rho}~,\end{equation} where $D$
 denotes maximal separation between the interferometer slits for
 which the interference still takes place for the light with wavelength $\bar{\lambda}$ received from a celestial
source located at a distance $r$ and having the size $\rho$. As we
stressed there is no effect in Eq.(\ref{explim}) due to
quantum-gravitational increment of $\bar{\lambda}$. Now by taking
the variations of $\rho,~r$ in Eq.(\ref{explim}) one
 finds
\begin{equation}\label{effrate} \delta D \simeq  D^{5/3} \left({l_P \over 0.16\cdot
    \bar{\lambda} \cdot r} \right)^{2/3}~. \end{equation} Let us
 estimate
the maximum of this variation by choosing the corresponding
parameters
 from
the data \cite{LH, RTG, Stein}, that is, $r \sim 1$kpc, $D \sim
 10^3$cm,
$\bar{\lambda} \sim 10^{-4}$cm. For this set of parameters from
 Eq.(\ref{effrate}) one finds

\begin{equation}\label{maxalphrate}  \delta D \sim
 \, 10^{-28}\mbox{cm}~. \end{equation}
The separation between the slits, $D$, for observations analyzed
in \cite{LH, RTG, Stein} varies from $1$m  to the $25$m. So that
the observations analyzed in \cite{LH, RTG, Stein} are simply
insensitive to such a small variation of $D$, that is, they have
no
 chance to detect the effect of quantum gravitational fluctuations.

\subsection*{ADD braneworld setup}

If $E_P \sim 10^{19}$GeV represents a proper quantum gravity
scale, then one can say at least two extremely different
fundamental scales, the electroweak scale $E_{EW}\sim 1$TeV and
the Planck scale $E_P$, appear to be present in the universe. The
fact that their ratio appears to be around $E_{EW}/E_P\sim
10^{-16}$ is a puzzle for many reasons. First, one can have
theoretical prejudice that a deeper comprehension of physics
should lead us to a theory with one single energy scale. So the
fact that gravity is so much weaker than other forces of Nature
seems a problem whose resolution will lead us to a better
understanding of our Universe. Second, even if we assume that the
fundamental theory has two different energy scales, one has to
understand what is there in the "desert" between these two scales,
and at which scale new physics will appear? This is a very
important question both for experimental purposes (is it worth
building accelerators to explore this desert?) and for theoretical
problems. In fact, the new physics scale is assumed to set the
ultraviolet cutoff for the presently known particle physics. It is
well known that the standard model of particle physics suffers
from a major theoretical problem, which is the stability of the
Higgs mass under radiative corrections: the Higgs mass is
quadratically sensitive to the ultraviolet cutoff and if the
cutoff scale is much higher than the electroweak scale an extreme
fine-tuning between the bare mass and the one-loop correction is
required to give a low value for the physical mass. It is
plausible therefore that the new physics scale to be very close to
$E_{EW}$. However the problem could still persist going up to the
Planck scale, which is the highest known scale, unless the new
physics is able to "screen" the sensitivity to $E_P$. This
possibility is the main motivation for models of low-scale
supersymmetry. However no hint for the low-scale supersymmetry has
been found in accelerators until now, and the arrival of the Large
Hadron Collider (LHC) calls for other possibilities. An
alternative possibility, attracting considerable continuing
interest proposed in the framework of models with large extra
dimensions assumes the presence of one fundamental scale and the
weakness of gravity comes from the fact that only gravity
propagates in the bulk \cite{ADD}. (For earlier braneworld
particle physics phenomenology one can see the papers
\cite{BarKanch}).

Let us briefly recapitulate the basics of ADD model. Extra
dimensions run from $0$ to $ L$ where the points $0$ and $ L$ are
identified \cite{ADD}. The standard model particles are localized
on the brane while the gravity is allowed to propagate throughout
the higher dimensional space and the fundamental scale of gravity
is taken to be close to the electroweak one, $E_F\sim$TeV. The
mass gap between the $n$th and $n+1$th KK modes is $\sim L^{-1}$
and correspondingly modification of Newton's inverse square law
(due to exchange of KK modes) takes place beneath the length scale
$L$. Roughly the gravitational potential on the brane produced by
the brane localized point-like particle $m$ looks like

\begin{equation}\label{potential}V(r)=\left\{\begin{array}{ll} l_F^{2+n}m/ r^{1+n}~, & \mbox{for}~~ r\lesssim L~,\\\\
l_{Pl}^2m/ r~, & \mbox{for}~~ r>L~.
\end{array}\right.\end{equation} From Eq.(\ref{potential}) one simply finds the relation between Planck
and fundamental lengths
\begin{equation}\label{relfundplanck}l_F^{2+n} \simeq L^nl_P^2~.\end{equation} Strictly
speaking the transition of four-dimensional gravity from the
region $r\gg L$ to the higher-dimensional law for $r\ll L$ is more
complicated near the transition scale $\sim L$ than it is
schematically described in Eq.(\ref{potential}), but it is less
significant for purposes of our discussion.

\subsection*{Operational definition of brane induced space-time}

{\bf \emph{Approach} 1. -} Let us repeat the discussions for
measurement of space-time distances by the brane localized clocks
and light signals. Nothing changes up to the
Eq.(\ref{minwitsize}). The upper bound on the mass of the clock is
set by the requirement the size of the clock not to be smaller
than its gravitational radius
\begin{equation}\label{uperbound}r_c^{min} \simeq \sqrt{{l\over
m_{max}}} \simeq r_g(m_{max})~,\end{equation} where $r_g$ denotes
gravitational radius of the clock. If the gravitational radius of
the clock is smaller than $L$, that is, $r_g < L$, trough the
Eq.(\ref{potential}) one finds \[r_g \simeq \left( l_F^{2+n}m
\right)^{1\over 1+n}~,\] and using this expression in
Eq.(\ref{uperbound}) the upper bound on the mass takes the form
\[m_{max} \simeq l^{1+n \over 3+n} l_F^{-{2(2+n)\over 3+n}}~.\]
Resorting back to the Eq.(\ref{minwitsize}) one estimates the
minimal uncertainty in length measurement as
\begin{equation}\label{HighKarol}\delta l_{min} \simeq l_F^{2+n\over 3+n}l^{1\over
3+n}~.\end{equation} If the gravitational radius of the clock is
greater than $L$, that is, $r_g > L$, one gets the
Eq.(\ref{Karol}).

{\bf \emph{Approach} 2. -} Nothing changes up to the
Eq.(\ref{eq:ineqtx}). If the fineness $\delta x$ is smaller than
$L$, that is $\delta x \lesssim L$, the requirement the clock not
to become black hole gives instead of Eq.(\ref{eq:gmx})

\begin{equation} \label{highsas} \left( l_F^{2+n}m \right)^{1\over 1+n} \lesssim \delta x~.
\end{equation} Combining Eqs.(\ref{highsas},\, \ref{eq:gmx}) the
Eq.(\ref{eq:txxx}) changes to \[ l_F^{2+n}t \lesssim \delta
x^{3+n}~, \] which by taking into account that our light-clock
having the size $\delta x$ can not measure the time to a better
accuracy than $\delta t = \delta x$ is nothing but the
Eq.(\ref{HighKarol}).

{\bf \emph{Approach} 3. -} If the size of the clock is smaller
than $L$, that is $r_c < L$, the gravitational time delay takes
the form
\[t'=\left(1-{2l_F^{2+n}m\over r_c^{1+n}}\right)^{1/2}t~.\] The disturbance of the
background metric to be small, the size of the clock should be
much greater than its gravitational radius $r_c \gg \left(
l_F^{2+n}m \right)^{1\over 1+n}$. Under this assumption for
gravitational disturbance in time measurement one finds
\[t'=\left(1-{l_F^{2+n}m\over r_c^{1+n}}\right)t~.\] Since we are using
light-clock its mass can not be less than $\pi/r_c$, which by
taking into account that the size of the clock determining its
resolution time represents in itself an error during the time
measurement gives \[\delta t \simeq  r_c + { l_F^{2+n}\,t\over
r_c^{2+n}}~,\] which after minimization with respect to $r_c$
leads to the Eq.(\ref{HighKarol}).

For the brane induced space-time also all these approaches lead to
the same result for space-time uncertainty, Eq.(\ref{HighKarol}),
but again one should notice that third approach strongly
discourages to take the optimal size of the clock to be close to
its gravitational radius. The first and second approaches do not
take into account the gravitational time delay of the clock and
correspondingly give the misleading results about the optimal
parameters of the clock. For the optimal parameters of the clock
for measuring the length scale $l \lesssim L^{3+n}l_F^{-(2+n)}$
one finds
\[r_c \simeq l_F^{2+n\over 3+n}\,l^{1\over 3+n}~,~~~~~~~~m \simeq {1\over
r_c}~.\] From these relations one easily finds that the
Eq.(\ref{HighKarol}) holds for the length scale \[l \lesssim
L^{3+n}l_F^{-(2+n)}~,\] which after using the relation
(\ref{relfundplanck}) (with $l_F \simeq 10^{-17}\mbox{cm}$) takes
the form \begin{equation}\label{validlebgthcon}l \lesssim
l_F^{3(2+n)\over n}l_P^{-{2(3+n)\over n}} \simeq 10^{96 + 15n\over
n}\mbox{cm}~.\end{equation}

\subsection*{Constraints on the braneworld scenarios}

Let us start with a simple example. Imprecision in length
measurement sets the limitation on the precision of energy
momentum measurement \[\lambda = 2\pi p^{-1}~~\Rightarrow ~~
\delta p = 2\pi\lambda^{-2}\delta\lambda~,~~ \delta E =
pE^{-1}\delta p~.\] The brane localized particle with momentum
grater than $L^{-1}$, probes the length scale beneath $L$ the
gravitational law for which is higher-dimensional. So, in this
case one can directly use the Eq.(\ref{HighKarol}) that gives
\begin{equation}\label{highenerunc}\delta p\sim {p^{1+\alpha}\over
E_F^{\alpha}}\,,~~~~~~\delta E\sim
{(E^2-m^2)^{{2+\alpha\over2}}\over E\,
E_F^{\alpha}}~,\end{equation} where $\alpha=(2+n)/(3+n)$. Using
this expression one can simply estimate that for ultra high energy
cosmic rays with
\[E\sim 10^{8}\mbox{TeV}~,\] the uncertainty in energy becomes
greater than \[ \delta E \simeq 10^{13}\mbox{TeV}~.\] The
experimental uncertainty of the energy of high-energy cosmic rays
is almost comparable to the energy itself, that is on the
experimental side we know \[\delta E \lesssim 10^8\mbox{TeV}~.\]
One simply finds that the ultra high energy cosmic rays put the
restriction on the fundamental scale \[E_F\gtrsim
10^{8}\mbox{TeV}~.\] From the GZK cutoff we know that the energy
of high energy cosmic proton drops below $10^{8}$TeV (through the
successive collisions on the typical CMBR photons accompanied by
the production of pions) almost independently upon initial energy
after it travels the distance of the order of $\sim 100$Mpc
\cite{GZK}. That is, protons detected with energies $ > 10^8$TeV
should be originated within the GZK distance $R_{\mbox{GZK}}
\simeq 100 $Mpc. But this mechanism is of little use against the
amplification of energy of the protons (coming usually from
distances greater than the GZK distance) through the background
metric fluctuations, Eq.(\ref{highenerunc}), as this amplification
takes place with equal probability within and outside of the GZK
distance. (In itself, as long as the energy scale of high energy
cosmic rays is much greater than the fundamental scale of gravity
their presence in theory needs a separate consideration
\cite{ADD2}).

Actually the situation is more dramatic. From
Eq.(\ref{highenerunc}) one sees that for the particle with the
mass $m\ll E_F$ and energy $E\sim E_F$, the uncertainty in energy
becomes comparable to the energy itself. So that the quantum
fluctuations of space-time become appreciable even for the TeV
scale physics.

Let us now estimate the effect on stellar interferometry
observations. The Eq.(\ref{HighKarol}) is valid beneath the length
scale given by Eq.(\ref{validlebgthcon}). With increasing the
number of extra dimensions this length scale decreases as: $n=2,\,
\sim 10^{63}\mbox{cm}\,;~n=3,\, \sim 10^{47}\mbox{cm}\,;~n=4,\,
\sim 10^{39}\mbox{cm}\,;~n=5,\, \sim 10^{34}\mbox{cm}\,;~n=6,\,
\sim 10^{31}\mbox{cm}\,;~n=7,\, \sim 10^{29}\mbox{cm}\,;~n=8,\,
\sim 10^{27}\mbox{cm}\,;~n=9,\, \sim 10^{26}\mbox{cm}\,;~n=10,\,
\sim 10^{25}\mbox{cm}$. For the increment of the wave packet width
one finds \[{\delta \omega \over
 \bar{\omega}} \simeq \,
10^{-13\alpha}~.\] By taking into account that in most
applications $r \gg \rho$ from Eq.(\ref{explim}) one finds
\[\delta D \simeq {0.16\, \bar{\lambda}\, r \,\delta \rho \over\rho^2} \simeq {l_F^{\alpha} D^{1+\alpha} \over
(0.16\,\bar{\lambda}\, r)^{\alpha} }~.\] For the set of parameters
$r \sim 1$kpc, $D \sim
 10^3$cm, $\bar{\lambda} \sim 10^{-4}$cm \cite{LH, RTG, Stein} one gets

\[\delta D \simeq 10^{3-29\alpha}\mbox{cm}~.\] In the case
$n = 2$ one gets $\delta \omega /
 \bar{\omega} \simeq \,
10^{-10}~\mbox{and}~\delta D \simeq 10^{-20}$cm, that is, in
comparison with Eq.(\ref{maxalphrate}) the effect is amplified by
$8$ orders of magnitude but still it is not so large to affect the
observations. So that stellar interferometry observations
considered in \cite{LH, RTG, Stein} are less sensitive to the
lowering of fundamental scale in the framework of large extra
dimensions.

From Eq.(\ref{highenerunc}) one sees that light speed is given
with the precision \begin{equation}\label{lighspeedvar}\delta
v_{group} = {d (\delta E) \over dp} \simeq \left({ E \over E_F}
\right)^{\alpha}~.\end{equation} Thus for photons emitted
simultaneously from a distant source coming towards our detector,
we expect an energy dependent spread in their arrival times. To
maximize the spread in arrival times, it is desirable to look for
energetic photons from distant sources. This proposal was first
made in another context in \cite{ACE}. The analyses of the TeV
flares observed from active galaxy Markarian 421 \cite{Gaid} puts
the limit on the variation of light speed with energy. This limit
applied to the Eq.(\ref{lighspeedvar}) gives the following
limitation on $E_F$ \cite{Bil, Sch}

\[   E_F \gtrsim 10^{16}\mbox{GeV}   ~. \]

All of the above restrictions are intimately related to the
modification of gravity Eq.(\ref{potential}) beneath the length
scale $L \gg l_P$. Therefore one can remove the above experimental
bounds in the case when gravity modification scale on the brane is
close to the length scale $\sim 10^{-30}$cm. But at the same time
we are interested to keep the fundamental scale of gravity, $E_F$,
close to the $E_{EW}$.

\subsection*{Shape modulus of extra space}

What can be a possible protecting mechanism from these
unacceptably amplified fluctuations for low lying fundamental
scale of gravity? Following the paper \cite{Di} let us take note
of the role of shape modulus of extra space. A flat,
two-dimensional toroidal compactification can be analyzed in much
details from this point of view \cite{Di}. Such a torus is
specified by three real parameters (the two radii $L_1,\,L_2$ of
the torus as well as the shift angle $\theta$), and corresponds to
identifying points which are related under the two coordinate
transformations

\begin{eqnarray}
  y_1 ~&\to&~~ y_1+2\pi L_1 \cos\theta~,\nonumber \\ y_2 ~&\to&~ ~y_2+2\pi L_2 \sin\theta \label{torusdef}~. \end{eqnarray}
Note that tori with different angles $\theta$ are topologically
distinct up to the modular transformations. While most previous
discussions of large extra dimensions have focused on the volume
of such tori essentially fixing $\theta=\pi/2$. Given the torus
identifications in Eq.~(\ref{torusdef}), it is straightforward to
determine the corresponding KK spectrum. The KK eigenfunctions for
such a torus are given by
\begin{equation}
     \exp\left\lbrack i {n_1\over L_1} \left(y_1- {y_2\over \tan\theta}\right)
                 ~+~ i{n_2\over L_2} {y_2\over \sin\theta} \right\rbrack~
\label{KKfuncts} \end{equation} where $n_i \in  Z$. Applying the
(mass)$^2$ operator
    $- (\partial^2 /\partial y_1^2 + \partial^2 /\partial y_2^2)$,
we thus obtain the corresponding KK masses
\begin{equation}
    M_{n_1,n_2}^2 ~=~ {1\over \sin^2\theta} \left(
         {n_1^2\over L_1^2} +
         {n_2^2\over L_2^2} - 2 {n_1n_2\over L_1L_2} \cos\theta\right)~.
\label{KKmasses} \end{equation} We see that while the KK spectrum
maintains its invariance under $(n_1,n_2)\to -(n_1,n_2)$, it is no
longer invariant under $n_1\to -n_1$ or $n_2\to -n_2$
individually.  The spectrum is, however, invariant under either of
these shifts and the simultaneous shift $\theta\to \pi-\theta$. We
can therefore restrict our attention to tori with angles in the
range $0<\theta\leq \pi/2$ without loss of generality. It is clear
from Eq.~(\ref{KKmasses}) that the KK masses depend on $\theta$ in
a non-trivial, level-dependent way. We are interested in the
behavior of the KK masses when the volume of the compactification
manifold is held fixed. For this purpose it is useful to
reparameterize the three torus moduli $(L_1,\,L_2,\,\theta)$ in
terms of a single real volume modulus $V$ and a complex shape
modulus $\tau$: \begin{equation}
    V \equiv 4\pi^2 L_1 L_2 \sin\theta~,~~~~~~~
    \tau \equiv {L_2\over L_1}\,e^{i\theta}~.
\label{moduli} \end{equation} We shall also define $\tau_1\equiv
{\rm Re}\,\tau$ and $\tau_2\equiv {\rm Im}\,\tau$. Using these
definitions, we can express $(L_1,\,L_2,\,\theta)$ in terms of
$(V,\tau)$ via \begin{eqnarray}
        && \cos\theta = \tau_1/|\tau| ~,~~~~ \sin\theta = \tau_2/|\tau|~,\nonumber\\
        && L_1^2 =  {1\over 4\pi^2 \tau_2}V~,~~~~
            L_2^2 =  {|\tau|^2 \over 4\pi^2 \tau_2}V~,
\end{eqnarray} that yields the
KK masses \begin{eqnarray}
       M_{n_1,n_2}^2 &=& {4\pi^2 \over V} {1\over \tau_2}
          \bigl| n_1\tau - n_2 \bigr|^2~\nonumber\\
         &=& {4\pi^2 \over V} {1\over \tau_2}
      \left\lbrack  (n_1\tau_1 - n_2)^2 + n_1^2 \tau_2^2 \right\rbrack~.
\label{KKmassestwo} \end{eqnarray} Note that although
Eq.~(\ref{KKmassestwo}) is merely a rewriting of
Eq.~(\ref{KKmasses}), we have now explicitly separated the effects
of the volume modulus $V$ from those of the shape modulus $\tau$.
At the expense of $\theta$ one can try to increase the mass gap
between KK modes with the fixed volume of extra space. One is
therefore led to study the limit $\theta\sim\epsilon \ll 1$
\begin{eqnarray}
    && \left( {V\over 4\pi^2 }\right) M_{n_1,n_2}^2 ~=~
             {(n_2-n_1 |\tau|)^2 \over |\tau| \epsilon}+\nonumber\\
     && ~~+ \left( {n_2^2 + 4 n_1 n_2 |\tau| + n_1^2 |\tau|^2 \over 6 |\tau|}
             \right) \epsilon +  {\cal O}(\epsilon^3)~.
\label{limitcasetwo} \end{eqnarray} (Note that in order to keep
the volume fixed as $\theta\to 0$, the radii are now forced to
grow increasingly large.) The first term on the right side of
Eq.~(\ref{limitcasetwo}) generally diverges when $|\tau|\equiv
L_2/L_1$ is irrational because in this case $n_2-n_1|\tau|$ never
vanishes exactly. Thus, the general KK state becomes infinitely
heavy as $\theta\to 0$. However, for any fixed chosen value of
$\epsilon$, we can always find special states $(n_1,n_2)$ for
which this first term comes {\it arbitrarily close}\/ to
cancelling;  this simply requires choosing sufficiently large
values of $(n_1,n_2)$.  These special states with large
$(n_1,n_2)$ are potentially massless. On the other hand, choosing
such large values of $(n_1,n_2)$ drives the second term in
Eq.~(\ref{limitcasetwo}) to larger and larger values. The third
and higher terms are always suppressed relative to the second term
in the $\epsilon\to 0$ limit, even as $(n_1,n_2)$ grow large. We
will not go into more analysis of the Eq.(\ref{limitcasetwo}) as
reader can find it in paper \cite{Di}, but simply indicate that in
certain cases (for small values of $\theta$) it is possible to
maintain the ratio between the higher-dimensional and
four-dimensional Planck scales while simultaneously increasing the
KK graviton mass gap by an arbitrarily large factor. This
mechanism can therefore be used to eliminate the above
experimental bounds on theories with large compact extra
dimensions.

\subsection*{Concluding remarks}

The way of reasoning presented in this paper is completely in the
spirit of quantum mechanics, that is to regard reality as that
which can be observed. First, following the discussions \cite{SW,
Mead, NvD, Sasakura, mazia, Karol}, we analyzed in a comparative
manner principal limitations on space-time measurement in light of
quantum mechanics and general relativity. All of the presented
approaches lead uniquely to the K\'arolyh\'azy uncertainty
relation (\ref{Karol}) but the third approach taking into account
the gravitational time delay reveals important disagreement
compared with the other approaches in estimating the optimal
parameters of the clock. Namely it tells us that optimal
parameters of the clock for measuring the space-time distance $l$
is given by \[r_c \simeq \delta l_{min}(l)\,,~~~~~~~~~~m \simeq {1
\over r_c}~,\] where $\delta l_{min}(l)$ denotes the uncertainty
in length measurement given by Eqs.(\ref{Karol},\,\ref{HighKarol})
in four and higher-dimensional scenarios respectively. Thus, from
Eq.(\ref{Karol}) one finds that for measuring the present Hubble
horizon $\sim 10^{28}$cm the optimal parameters of the clock are
estimated as $r_c \simeq 10^{-13}$cm,\,\,$m \simeq 1$GeV.
Hitherto, say in the framework of approaches {\bf \emph 1} and
{\bf \emph 2}, it was understood mistakenly that the size of an
optimal clock had to be close to its gravitational radius, that
is, the mass of such a clock was defined as $m = r_c/l^2_P$. The
reason of this misconception was the disregard of gravitational
time delay.

Operational definition of space-time in light of quantum mechanics
and general relativity indicates an expected imprecision in
space-time structure. The resultant intrinsic imprecision in
space-time structure is quantified by the K\'arolyh\'azy
uncertainty relation. This relation sheds new light on the
relation between IR and UV scales in effective quantum field
theory satisfying black hole entropy bound \cite{CKN}. In spite of
the fact that minimal uncertainty in distance measurement given by
the K\'arolyh\'azy uncertainty relation is much greater than the
Planck length (provided $l \gg l_P$), the rate of
quantum-gravitational fluctuations is still controlled by the
Planck scale and is therefore discouragingly small to be
detectable by the present experiments and observations.
Nevertheless the rate of fluctuations can become unacceptably
amplified when the fundamental scale of gravity is lowered in the
framework of large extra dimensions. It is important to notice
that this amplification of fluctuations is not directly related to
the low quantum gravity scale but rather to the higher-dimensional
modification of Newton's law at relatively large distances.
Therefore the models with compact extra dimensions can be
protected from these fluctuations at the expense of shape modulus
of extra space. That is, we can keep the volume of extra space
fixed in order to have the low fundamental scale of gravity (see
relations (\ref{relfundplanck}) and (\ref{moduli})) but at the
same time using the shape modulus of extra space we can enlarge
the mass gap between KK modes to reduce the length scale at which
the modification of Newton's inverse square law takes place
\cite{Di}. This procedure can remove the above experimental bounds
on the fundamental scale of gravity as they arise because of
relatively large length scale at which Newton's inverse square law
of gravity changes to the higher-dimensional one. Presented
considerations demonstrate dramatic difference between braneworld
models with compact and open extra dimensions respectively. The
models with low fundamental scale of gravity having open extra
dimensions may be in serious trouble as there seems almost no
natural way to protect them from the unacceptably amplified
quantum-gravitational fluctuations.

\begin{acknowledgments}

Author is greatly indebted to Professors Jean-Marie Fr\`{e}re and
Peter Tinyakov for invitation and hospitality at the \emph{Service
de Physique Th\'eorique,
  Universit\'e Libre de Bruxelles}, where this paper was started, and to
Professor Seifallah Randjbar-Daemi for invitation to the
\emph{Abdus Salam International Centre for Theoretical Physics},
where this paper was finished. It is a pleasure to acknowledge
useful discussions with G.~Dvali, N.~Grandi, T.~Kahniashvili,
G.~Senjanovic and P.~Tinyakov. The work was supported by the
\emph{INTAS Fellowship for Young Scientists} and the
\emph{Georgian President Fellowship for Young Scientists}.

\end{acknowledgments}

\end{document}